\documentclass{article}%
\usepackage{amsmath}
\usepackage{amsfonts}
\usepackage{amssymb}
\usepackage{graphicx}%
\newcommand{\bpartial}{\mathop{\partial\kern -4pt\raisebox{.8pt}{$|$}}}
\newcommand{\bra}{\mathopen{[\kern-1.6pt[}}
\newcommand{\ket}{\mathclose{]\kern-1.5pt]}}
\newcommand{\bbra}{\mathopen{[\kern-2.2pt[\kern-2.3pt[}}
\newcommand{\bket}{\mathclose{]\kern-2.1pt]\kern-2.3pt]}}

\makeindex
\begin{document}

\title {\large{ \bf Exact three dimensional black hole with gauge fields in string theory }}

\vspace{3mm}

\author {  \small{ \bf S. Hoseinzadeh }\hspace{-2mm}{ \footnote{ e-mail: hoseinzadeh@azaruniv.edu}} { \small
and}
\small{ \bf A. Rezaei-Aghdam }\hspace{-2mm}{ \footnote{Corresponding author. e-mail:
rezaei-a@azaruniv.edu}} \\
{\small{\em Department of Physics, Faculty of Science, Azarbaijan Shahid Madani University, }}\\
{\small{\em  53714-161, Tabriz, Iran  }}}

\maketitle

\begin{abstract}

We have obtained exact three dimensional $BTZ$ type solutions
with gauge fields, for string theory on a gauge symmetric
gravitational background constructed from semi-simple extension
of the Poincar\'{e} algebra (and the Maxwell algebra) in $2+1$
dimensions. We have
studied the models for two non-Abelian and Abelian gauge fields
solutions and shown that the related sigma models for each of
these backgrounds is a $SL(2,R)$ WZW (Wess-Zumino-Witten) model
and that they are classically canonically equivalent. We have
also obtained the dual solution for the Abelian case and by
interpreting the new field strength tensors of the Abelian
solution as electromagnetic field strength tensors shown that
dual models coincide with the charged black string solution.

\end{abstract}
\newpage
\section {\large {\bf Introduction}}
\setlength{\parindent}{0cm}

In order to overcome some complexities of four dimensional
gravity, many of researchers have studied the gravity models in
lower dimensions. For instance, they hope that some properties of lower
dimensional black holes help them to model those of the four
dimensional black holes. One of such attempts has resulted in the
construction of the three dimensional rotating BTZ black hole
\cite{Banados}. Nearly two decade ago, Witten \cite{Witten} has
shown that an exact two dimensional black hole in string theory
could be obtained by gauging a one dimensional subgroup $U(1)$ of
$SL(2,R)$. Also, exact three dimensional black string \cite{Horne}
and black hole \cite{Horowitz} solutions in string theory have
been obtained. In other attempt, the $SL(2,R)$ WZW model and it's relation
to string theory in $AdS_{3}$ has been studied in details in three
different papers \cite{Maldacena-1,Maldacena-2,Maldacena-3}. In those papers, the structure of the Hilbert
space of the WZW model and the spectrum of physical states of the
string theory have been determined \cite{Maldacena-1}, and also
the one loop amplitude \cite{Maldacena-2} and the correlation
functions of the model \cite{Maldacena-3} have been studied.

Recently, the Maxwell \cite{Bacry,Schrader} and Semi-simple
extension of Poincar\'{e} symmetries was applied to construct
gauge symmetric gravity models in $3+1$ \cite{Azcarraga,Soroka}
and $2+1$ \cite{Hoseinzadeh} dimensions (see also \cite{Salgado, Diaz}). Here, we apply these
symmetries to obtain an exact three dimensional black hole with
gauge fields in string theory by introducing a new extended
anti-symmetric B-field and try to obtain some solutions for
equations of motion of the low energy string effective action
\cite{Tseytlin,Callan}. The outlines of the paper is
as follows:

In section two, we construct a low energy string effective action
in $2+1$ dimensions by use of new gauge field strengths and
obtain it's equations of motion. In section three, we solve the
equations of motion using the $BTZ$ metric and obtain two
different non-Abelian and Abelian solutions (Abelian solution has
no contribution of interaction terms in new gauge field
strengths). Then, we show that both solutions are {\it exact
solutions} whereas the sigma models for each of these backgrounds
is a $SL(2,R)$ Wess-Zumino-Witten model. We also show that two
sigma models corresponding to the two non-Abelian and Abelian
solutions are classically canonically equivalent. We interpret
the new gauge field strengths of the Abelian solution as an
electromagnetic field strength tensors, and obtain the
corresponding three different electric and magnetic fields. In
section four, using duality transformation, we calculate dual
solutions of the Abelian solution with respect to both spacelike
($\varphi$) and timelike coordinate isometry symmetries and show
that the dual solutions correspond to the charged black string
solution. By obtaining both $\varphi$-dual and $t$-dual electric
and magnetic fields, we show that duality relates the electric
fields to the dual magnetic fields and vice versa. We present some
concluding remarks in section five. In appendix, we cast both the
Maxwell algebra and the semi-simple extension of the Poincar\'{e}
algebra (Ads-Lorentz algebra\footnote{The semi-simple extension
of the Poincar\'{e} algebra is the direct sum of the $Ads$
algebra and the Lorentz algebra, i.e. $so(2,2)\oplus so(2,1)$ in
$2+1$ dimensional spacetime, and then is called the Ads-Lorentz
algebra.}) in $2+1$ dimensional spacetime from the anti-de Sitter
(Ads) algebra (so(2,2) in $2+1$ dimensions) by use of the
S-expansion procedure with the appropriate semigroups $S$ and
$\overline{S}$ \cite{Izaurieta-1,Izaurieta-2}.

\section {\large {\bf Gauge fields in string action from semi-simple extension of the
Poincar\'{e} algebra in $2+1$ dimensions}}
\setlength{\parindent}{0cm}

Let us consider the semi-simple extension of the Poincar\'{e}
algebra (or the Maxwell algebra for $\lambda=0$) with the basis
$X_{B}=\{P_{a},J_{a},Z_{a}\}$ in $2+1$ dimensions
\cite{Hoseinzadeh} as follows:\footnote{The relationship between
Maxwell algebra and semi-simple extension of the Poincar\'{e}
algebra with the $Ads$ algebra has been discussed in the
appendix.}
\begin{equation}   \nonumber
[J_{a},J_{b}] = \epsilon_{abc} J^{c}, ~~~~~~~[J_{a},P_{b}] =
\epsilon_{abc} P^{c}, ~~~~~~~~~~[P_{a},P_{b}] = k \epsilon_{abc}
Z^{c},~~~
\end{equation}
\begin{equation} \label{L1}
[J_{a},Z_{b}] = \epsilon_{abc} Z^{c}, ~~~~~~[P_{a},Z_{b}] =
-\frac{\lambda}{k} \epsilon_{abc} P^{c}, ~~~~~~[Z_{a},Z_{b}] =
-\frac{\lambda}{k} \epsilon_{abc} Z^{c},
\end{equation}
where $k$ and $\lambda$ are constants, and $Z_{a}$ are new
generators which are added to ordinary Poincar\'{e} generators
$P_{a}$ and $J_{a}$ to extend the algebra. One can construct gauge
fields $h_{\mu}$ which are Lie algebra valued one form $h=h_{\mu}
dx^{\mu}$ as follows:
\begin{equation}\label{L2}
h_{\mu}(x)=h_{\mu}^{~B}(x)X_{B}=e_{\mu}^{~a}(x)P_{a}+\omega_{\mu}^{~a}(x)J_{a}+A_{\mu}^{~a}(x)Z_{a},
\end{equation}
where $x^{\mu}~(\mu=0,1,2)$ and $a,b=0,1,2$ are spacetime
coordinates and Lie algebra indices, respectively. Using these
gauge fields $\Big(e_{\mu}^{~a}(x), \omega_{\mu}^{~a}(x),
A_{\mu}^{~a}(x)\Big)$, one can write the new field strengths
$F^{~~a}_{\mu\nu}$ as follows \cite{Hoseinzadeh}:
\begin{equation}\label{L3}
F_{\mu \nu}^{~~a}=\partial_{[\mu} A_{\nu]}^{~a} +
\epsilon_{bc}^{~~a}(k \: e_{\mu}^{\ b} e_{\nu}^{\ c} +
\omega_{\mu}^{\ b} A_{\nu}^{\ c}+A_{\mu}^{\ b} \omega_{\nu}^{\
c}) -\frac{\lambda}{k} \epsilon_{bc}^{~~a} A_{\mu}^{~b} \;
A_{\nu}^{~c}.
\end{equation}
where the gauge fields $e_{\mu}^{~a}$, $\omega_{\mu}^{~a}$ and
$A_{\mu}^{~a}$ are vierbein, spin connection and new non-Abelian
gauge fields\footnote{Note that for the general form of the field
strength \eqref{L3} the gauge fields $A_{\mu}^{a}$ are
non-Abelian. For the Abelian case we have:
\begin{equation}\nonumber
F_{\mu \nu}^{~~a}=\partial_{\mu} A_{\nu}^{~a} - \partial_{\nu}
A_{\mu}^{~a},
\end{equation}
i.e. for these gauge fields $A_{\mu}^{~a}$ we have:
\begin{equation}\nonumber
\epsilon_{bc}^{~~a}(k \: e_{\mu}^{\ b} e_{\nu}^{\ c} +
\omega_{\mu}^{\ b} A_{\nu}^{\ c}+A_{\mu}^{\ b} \omega_{\nu}^{\
c}) -\frac{\lambda}{k} \epsilon_{bc}^{~~a} A_{\mu}^{~b} \;
A_{\nu}^{~c}=0.
\end{equation}}
(corresponding to the new generators $Z_{a}$) respectively. Now we
consider the low energy string effective action in $2+1$
dimensional spacetime $\mathcal{M}$ by use of these gauge fields
as follows: \footnote{Note that the above action is the ordinary
string effective action in $2+1$ dimensions where the last term
is added.} \cite{Horowitz, Tseytlin}
\begin{equation}\label{L4}
S=\int_{\mathcal{M}} d^{3}x \sqrt{-g}~e^{-2\phi} \Big[
R+\frac{4}{K}+4(\nabla \phi)^2 -\frac{1}{12}H_{\mu\nu\rho}
H^{\mu\nu\rho} -\frac{1}{12}
 H^{\prime~~~a}_{\mu\nu\rho} H^{\prime\mu\nu\rho}_{~~~~a}   \Big]
\end{equation}
where $R$ is the Ricci scalar of $\mathcal{M}$ and $\phi$ is the
dilaton field, and furthermore $H_{\mu\nu\rho}$ and
$H^{\prime~~~a}_{\mu\nu\rho}$ \footnote{The algebra indices $a$ can be taken up and down by the ad-invariant metric $\Omega_{ab}$ of the algebra \cite{Hoseinzadeh}.} are defined as follows:
\begin{equation}\label{L5}
H_{\mu\nu\rho}=\partial_{\mu} B_{\nu\rho}+\partial_{\nu}
B_{\rho\mu}+\partial_{\rho} B_{\mu\nu},~
\end{equation}
\begin{equation}\label{L6}
H^{\prime ~~~a}_{\mu\nu\rho}=\partial_{\mu}
F^{~~a}_{\nu\rho}+\partial_{\nu}
F^{~~a}_{\rho\mu}+\partial_{\rho} F^{~~a}_{\mu\nu},
\end{equation}
so that $B_{\mu\nu}$ is an antisymmetric field and
$F^{~~a}_{\mu\nu}$ are the new antisymmetric field strengths.
Now, one can find the following equations of motion (the beta
functions) by variations of the above action with respect to
$g^{\mu\nu}$, $B^{\nu\rho}$, $F_{~~a}^{\nu\rho}$ and $\phi$
respectively,
\begin{equation} \nonumber
~~~~~~~~~~~~R_{\mu\nu}+2\nabla_{\mu}\nabla_{\nu}\phi-\frac{1}{4}H_{\mu\rho\sigma}H_{\nu}^{~\rho\sigma}
-\frac{1}{8}H_{\mu\rho\sigma}^{\prime ~~~a}H_{\nu~~~a}^{\prime~
\rho\sigma}=0,
\end{equation}
\begin{equation}\nonumber
~~~~~~~~~~~~~~~~~~~~~~~~~~~~~~~~~~~~~~~~~~~~~~~~~~~~~\nabla^{\mu}(e^{-2\phi}
H_{\mu\nu\rho})=0,
\end{equation}
\begin{equation}\label{L7}
~~~~~~~~~~~~~~~~~~~~~~~~~~~~~~~~~~~~~~~~~~~~~~~~~~~~\nabla^{\mu}(e^{-2\phi}
H_{\mu\nu\rho}^{\prime ~~~a})=0,
\end{equation}
\begin{equation} \nonumber
4\nabla^{2}\phi-4(\nabla\phi)^{2}+R+\frac{4}{K}-\frac{1}{12}H_{\mu\nu\rho}
H^{\mu\nu\rho} -\frac{1}{12}  H^{\prime~~~a}_{\mu\nu\rho}
H^{\prime\mu\nu\rho}_{~~~~a}=0,~~~~~~~~~~
\end{equation}
where these equations are zeros of the beta functions (at one
loop) $\beta(G)$, $\beta(B)$, $\beta(F^{a})$ and $\beta(\phi)$
(respectively) for the following sigma model \cite{Callan}
\begin{equation}\label{L8}
I=\int_{\Sigma} d^{2}\sigma \sqrt{g} \Big(G_{\mu\nu}(X)
\partial_{\alpha} X^{\mu} \partial_{\beta} X^{\nu} g^{\alpha\beta}
+B^{\prime}_{\mu\nu}(X) \partial_{\alpha} X^{\mu}
\partial_{\beta} X^{\nu} \epsilon^{\alpha\beta} +\frac{1}{2} R^{(2)} \phi(X) \Big),
\end{equation}
where $\Sigma$ is the worldsheet, $R^{(2)}$ is the scalar curvature of the worldsheet metric $g_{\alpha\beta}$, and $\epsilon^{\alpha\beta}$ is
an anti-symmetric 2-tensor, normalized so that $\sqrt{g}
\epsilon^{12}=+1$, and the extended antisymmetric B-field has the
following form:
\begin{equation}\label{L9}
B^{\prime}_{\mu\nu}(X) =B_{\mu\nu}(X) +F^{~0}_{\mu\nu}(X)
+F^{~1}_{\mu\nu}(X) +F^{~2}_{\mu\nu}(X) =B_{\mu\nu}(X)
+F^{~a}_{\mu\nu}(X)~\xi_{a},
\end{equation}
such that we have $\xi_{a}=(1,1,1)$.

\section {\large {\bf Black hole solutions }}
\setlength{\parindent}{0cm}

In this section, we will try to obtain a solution for the
equations \eqref{L7}. We know that the $2+1$ dimensional
Einstein-Hilbert action with cosmological constant term
\begin{equation}\label{L10}
S=\int d^{3}x \sqrt{-g} \Big( R-2\Lambda  \Big),
\end{equation}
has a BTZ black hole solution as follows: \cite{Banados}
\begin{equation}\label{L11}
ds^2=-N^2(r) dt^2+ \frac{1}{N^2(r)} dr^2 +r^2 (N^{\phi}(r) ~ dt + d \phi)^2,
\end{equation}
with
\begin{equation}\label{L12}
N^2(r) = -M + \frac{r^{2}}{\ell^{2}}+ \frac{J^{2}}{4 r^2} ,~~~~~~~~~~~~~~   N^{\phi}(r) =- \frac{J}{2 r^2},
\end{equation}
where $\Lambda=-\frac{1}{\ell^{2}}$ is
the negative cosmological constant, furthermore $M$ and $J$ are
the mass and angular momentum of the black hole, respectively.
Here $[x^{0}, x^{1}, x^{2}] = [t, r, \varphi]$ are the
coordinates of the spacetime. We use the above BTZ metric to
solve the equations of motion \eqref{L7} assuming that all fields
are a function of the radial coordinate only. Here, we analyse
two interesting non-Abelian and Abelian gauge field solutions.

\subsection {\large {\bf Non-Abelian case }}

Using the form of the metric as \eqref{L11} and \eqref{L12} the
equations of motion \eqref{L7} have a solution as follows:
\begin{equation}\nonumber
B_{20}(r)=\frac{r^2}{\ell},~~~~~~~~~~ \phi(r)=0,~~~~~~~~~~ K=\ell^{2},~~~~~~~~~~~~~~~~~~~~~~~~~~~~~~~
\end{equation}
\begin{equation}\nonumber
\omega^{0}(r)=\frac{v(r)}{D_{2}r^2+D_{1}} \Big(\pm r^2
\sqrt{D_{2}^{~2}-D_{4}^{~2}}-krN(r)+D_{5}\Big)dt +u(r)dr,
\end{equation}
\begin{equation}\label{L13}
\omega^{1}(r)=w(r)dr,~~~~~~~~~\omega^{2}(r)=v(r)dt+z(r)dr,~~~~~~~~~~~~~~~~~~~~~~~~~~
\end{equation}
\begin{equation}\nonumber
A^{0}(r)=\frac{D_{4}r^2+D_{3}}{v(r)}\Big(-\frac{J}{2r^2}dt+d\varphi\Big)+y(r)dr,~~~~~~~~~~~~~~~~~~~~~~~~~~~~
\end{equation}
\begin{equation}\nonumber
A^{1}(r)=\frac{D_{2}r^2+D_{1}}{v(r)}\Big(-\frac{J}{2r^2}dt+d\varphi\Big)+q(r)dr,~~~~~~~~ A^{2}(r)=s(r)dr,
\end{equation}
where $D_{1},D_{2},D_{3},D_{4},D_{5}$ are arbitrary constants and
$v(r)\neq0,u(r),w(r),y(r),q(r),s(r),z(r)$
are arbitrary functions of radial coordinate $r$ only. Now, using
\eqref{L3}, we find the following nonzero components of the new antisymmetric
fields \eqref{L3} for the non-Abelian gauge fields $A_{\mu}^{~a}$:
\begin{equation}\nonumber
F^{~~0}_{20}(r)=D_{2}r^2+D_{1},~~~~~~~~~~~~
\end{equation}
\begin{equation}\label{L14}
F^{~~1}_{20}(r)=D_{4}r^2+D_{3},~~~~~~~~~~~~
\end{equation}
\begin{equation}\nonumber
F^{~~2}_{20}(r)=r^2\sqrt{D_{2}^{~2}-D_{4}^{~2}}+D_{5},
\end{equation}
which yields the following extended antisymmetric B-field:
\begin{equation} \nonumber
B^{\prime}_{20}(r) =B_{20}(r) +F^{~~0}_{20}(r) +F^{~~1}_{20}(r)
+F^{~~2}_{20}(r)~~~~~~~~~~~~~~~~~~~~~~~~~~~~~
\end{equation}
\begin{equation} \label{L15}
=\Big(\frac{1}{\ell}+D_{2}+D_{4}+\sqrt{D_{2}^{~2}-D_{4}^{~2}}\Big)~r^2+D_{1}+D_{3}+D_{5}.
\end{equation}
Although this is a solution for the one-loop beta function
equations, one can show that by selecting
$D_{2}=D_{4}=-\frac{1}{\ell}$, this solution is also an {\it exact
solution} of the beta function equations in all loops. Indeed, the sigma model
\eqref{L8} with this background $(G_{\mu\nu},
B^{\prime}_{\mu\nu})$ is a $SL(2,R)$ Wess-Zumino-Witten model.
One can easily check this, by using the following $SL(2,R)$ group
element:
\begin{equation}\label{L16}
g(t,r,\varphi) =\left( \begin{tabular}{cc}
                  $ \hat{r} ~e^{-\hat{\varphi}} $     &    $\sqrt{\ell^{2}-\hat{r}^{2}}~e^{\frac{\hat{t}}{\ell}}$      \\
                  $\sqrt{\ell^{2}-\hat{r}^{2}}~e^{-\frac{\hat{t}}{\ell}}$     &   $ -\hat{r} ~e^{\hat{\varphi}} $       \\
                \end{tabular} \right),
\end{equation}
in the $WZW$ action
\begin{equation}\nonumber
S_{WZW}=\frac{\overline{k}}{4\pi} \int d^{2}z~ Tr(g^{-1} \partial
g g^{-1} \overline{\partial} g) -\frac{\overline{k}}{12\pi} \int
Tr(g^{-1} d g)^{3},
\end{equation}
where
\begin{equation}\label{L17}
\hat{r}=\ell
\sqrt{\frac{r^{2}-r_{-}^{2}}{r_{+}^{2}-r_{-}^{2}}},~~~~~~~~~~~~~~~
\hat{t}=\frac{r_{+}}{\ell}t+r_{-}\varphi,~~~~~~~~~~~~~~
\hat{\varphi}=\frac{r_{-}}{\ell^{2}}t+\frac{r_{+}}{\ell}\varphi,
\end{equation}
so that $\overline{k}$ is level of the $WZW$ model, and the horizons
$r=r_{\pm}$ of the black hole have the following relation to it's
mass and angular momentum $(M,J)$
\begin{equation}\label{L18}
r_{\pm}=\frac{\ell}{2}\Big(\sqrt{M+\frac{J}{\ell}}\pm\sqrt{M-\frac{J}{\ell}}~\Big).
\end{equation}

\subsection {\large {\bf Abelian (electromagnetic) case }}

Using the form of the metric as \eqref{L11} and \eqref{L12}, we
find another solution for the equations of motion \eqref{L7} as
follows:
\begin{equation}\nonumber
B_{20}(r)=\frac{r^2}{\ell},~~~~~~~~~~~~~ \phi(r)=0,~~~~~~~~~~~~
K=\ell^{2},~~~~~~~~~~~~~~~~~~~~~~~~~~~~~~~~~~~~~~~~
\end{equation}
\begin{equation}\nonumber
\omega^{0}(r)=\frac{g(r)}{r} N(r)~dt +
\Big(\frac{-f(r)h(r)g(r)+p(r)(\lambda
r^2+(g(r))^2)}{kr^2}\Big)~dr,~~~~~~~~~~~~~~~
\end{equation}
\begin{equation}\label{L19}
\omega^{1}(r)=-\frac{J}{2r^2}g(r)dt +f(r)dr +
g(r)d\varphi,~~~~~~~~~~ \omega^{2}(r)=\frac{-\lambda
r}{g(r)N(r)}~dr,~~~~~~~~~~~
\end{equation}
\begin{equation}\nonumber
A^{0}(r)=-\frac{J}{2r^2}h(r)dt + p(r)dr +
h(r)d\varphi,~~~~~~~~~~~~~~~~~~~~~~~~~~~~~~~~~~~~~~~~~~~~~~~~~~
\end{equation}
\begin{equation}\nonumber
A^{1}(r)=\frac{Jk}{2 g(r)}dt - \frac{k r^2}{g(r)}d\varphi,~~~~~~~~~~~~
A^{2}(r)=\frac{-kr}{g(r)N(r)}dr,~~~~~~~~~~~~~~~~~~~~~~~~~~~
\end{equation}
where $g(r)\neq0,f(r),h(r),p(r)$ are arbitrary functions of
radial coordinate only. Note that for this solution, $A_{\mu}^{~a}$ is Abalian and hence yields zero contributions
for the coupling terms in all components of the new field
strengths $F^{~~a}_{\mu\nu}$ \Big(i.e. we have ~$\epsilon_{bc}^{~~a}(k \: e_{\mu}^{\ b} e_{\nu}^{\ c} + \omega_{\mu}^{\ b} A_{\nu}^{\ c}+A_{\mu}^{\ b} \omega_{\nu}^{\ c}) -\frac{\lambda}{k} \epsilon_{bc}^{~~a} A_{\mu}^{~b} \; A_{\nu}^{~c}=0$\Big), and therefore
$F^{~~a}_{\mu\nu}$ can be interpreted as electromagnetic
field strength tensors without coupling terms as follows:
\begin{equation}\label{L20}
F_{\mu \nu}^{~~a}=\partial_{\mu} A_{\nu}^{~a} - \partial_{\nu} A_{\mu}^{~a}.
\end{equation}
For the above solution, all of the nonzero components of the
electromagnetic field strength tensors are as follows:
\begin{equation} \nonumber
F^{~~0}_{01}(r)=-\frac{J}{r^3}h(r)+\frac{J}{2r^2}\frac{d}{dr}h(r),~~~~~~~~~~~~~
F^{~~0}_{21}(r)=-\frac{d}{dr}h(r),~~~~~~~~~~~~~~~~~
\end{equation}
\begin{equation}\label{L21}
F^{~~1}_{01}(r)=\frac{Jk}{2
g^2(r)}\frac{d}{dr}g(r),~~~~~~~~~~~~~~~~~~~~~~~~~
F^{~~1}_{21}(r)=\frac{2kr}{g(r)}-\frac{k
r^2}{g^2(r)}\frac{d}{dr}g(r).~~~
\end{equation}
In this way, the extended B-field components have the following forms:
\begin{equation} \nonumber
B^{\prime}_{20}(r)
=B_{20}(r)=\frac{r^{2}}{\ell},~~~~~~~~~~~~~~~~~~~~~~~~~~~~~~~~~~~~~~~~~~~~~~~~~~~~~~~~~~~~~~~~~~~~~~~~~~~~~~~~~~~~~~
\end{equation}
\begin{equation} \label{L22}
B^{\prime}_{01}(r) =B_{01}(r) +F^{~~0}_{01}(r) +F^{~~1}_{01}(r)
+F^{~~2}_{01}(r)
=-\frac{J}{r^3}h(r)+\frac{J}{2r^2}\frac{d}{dr}h(r)+\frac{Jk}{2
g^2(r)}\frac{d}{dr}g(r),
\end{equation}
\begin{equation} \nonumber
B^{\prime}_{21}(r) =B_{21}(r) +F^{~~0}_{21}(r) +F^{~~1}_{21}(r)
+F^{~~2}_{21}(r)
=-\frac{d}{dr}h(r)+\frac{2kr}{g(r)}-\frac{k
r^2}{g^2(r)}\frac{d}{dr}g(r).~~~~~~~~~~
\end{equation}
where we assume that $B_{01}(r)=B_{21}(r)=0$. This solution is
also an exact solution. One can show that using the
$SL(2,R)$ group element \eqref{L16} with
\begin{equation}\label{L23}
\hat{r}=\ell
\sqrt{\frac{r^{2}-r_{-}^{2}}{r_{+}^{2}-r_{-}^{2}}},~~~~~~~~~~~~~~~
\hat{t}=\frac{r_{+}}{\ell}t-r_{-}\varphi,~~~~~~~~~~~~~~~~
\hat{\varphi}=\frac{r_{-}}{\ell^{2}}t-\frac{r_{+}}{\ell}\varphi,
\end{equation}
the sigma model \eqref{L8} with the background $(G_{\mu\nu},
B^{\prime}_{\mu\nu})$ is a $SL(2,R)$ $WZW$ model.
Now, we show that two sigma models corresponding to the two
discussed non-Abelian \eqref{L13} and Abelian \eqref{L19}
solutions are classically canonically equivalent. Indeed, by assuming the
following relations among the arbitrary functions in \eqref{L13} and
\eqref{L19},
\begin{equation}\nonumber
v(r)= \frac{2r^2 \sqrt{-\Lambda}}{h(r)-\frac{k
r^2}{g(r)}-\int dr\frac{k
r}{N(r)}+D}~,~~~~~~~~~~~~~~~~~~~~~~~~~~~~~~~~~~~~~~~~~~~~~
\end{equation}
\begin{equation}\label{L24}
y(r)= \Big(\frac{Jk}{r g(r)}-\frac{J
h(r)}{r^3}\Big)\frac{1}{v(r)}+\frac{2J \sqrt{-\Lambda}}{r
v^2(r)}-q(r)+\frac{k N(r)}{r
\sqrt{-\Lambda}}\Big(s(r)-q(r)\Big),
\end{equation}
\begin{equation}\nonumber
u(r)= w(r)-2 z(r) + \frac{\lambda}{k}~\Big(y(r) -q(r) +2
s(r)\Big),~~~~~~~~~~~~~~~~~~~~~~~~~~~~~~~~
\end{equation}
then, these two sigma models can be canonically related to each other by the
following relations \cite{Sfetsos}
\begin{equation}\nonumber
(C^{-1})^{\mu\rho}(G_{\rho\lambda}+B^{\prime}_{\rho\lambda}+A_{\rho\lambda})
(C^{-1})^{\kappa\lambda}(\widetilde{G}_{\kappa\nu}+\widetilde{B}^{\prime}_{\kappa\nu}+\widetilde{A}_{\kappa\nu})
=\delta^{\mu}_{~\nu},
\end{equation}
\begin{equation}\label{L25}
(G_{\mu\nu}+B^{\prime}_{\mu\nu})=(G_{\mu\rho}+B^{\prime}_{\mu\rho}+A_{\mu\rho})
(C^{-1})^{\lambda\rho}(\widetilde{G}_{\lambda\nu}+\widetilde{B}^{\prime}_{\lambda\nu}),
\end{equation}
where $(G_{\mu\nu},B^{\prime}_{\mu\nu})$ and
$(\widetilde{G}_{\mu\nu},\widetilde{B}^{\prime}_{\mu\nu})$ are the
non-Abelian \eqref{L13} and the Abelian \eqref{L19} solutions
respectively, such that $C_{\mu\nu}, A_{\mu\nu}$ and
$\widetilde{A}_{\mu\nu}$ have the following forms
\begin{equation}\label{L26}
C_{\mu\nu}=G_{\mu\nu}+B^{\prime}_{\mu\nu},~~~~~~~~~~~~
A_{\mu\nu}=\widetilde{A}_{\mu\nu}=0.
\end{equation}
Now, for the Abelian solution, the electric and magnetic fields can be written in terms of
electromagnetic field strength tensors as follows:
\begin{equation}\nonumber
E_{r}^{(a)}=-F^{01a}=-(g^{00}g^{11}F_{01}^{~~a}+g^{02}g^{11}F_{21}^{~~a})
=F_{01}^{~~a}-N^{\phi}(r)F_{21}^{~~a},~~~~~~~~~~
E_{\varphi}^{(a)}=0,~~~~
\end{equation}
\begin{equation}\label{L27}
B_{z}^{(a)}=rF^{12a}=-r(g^{11}g^{20}F_{01}^{~~a}+g^{11}g^{22}F_{21}^{~~a})
=\frac{J}{2r}F_{01}^{~~a}-(\frac{N^2-r^2(N^{\phi})^2}{r})F_{21}^{~~a},~~~~~~~~~~
\end{equation}
where the indices inside the parenthesis are algebra indices
which run over 0,1,2 and each denote a different field. In this
way, we have {\it three different electric and magnetic fields}. The
radial components of three electric fields have the following
forms:
\begin{equation}\label{L28}
E_{r}^{(0)}=-\frac{J}{r^3}h(r),~~~~~~~~~~ E_{r}^{(1)}=\frac{Jk}{r
g(r)},~~~~~~~~~~ E_{r}^{(2)}=0,~~~~~~~~~
\end{equation}
furthermore, all of the azimuthal components of them are zero
\begin{equation}\label{L29}
E_{\varphi}^{(0)}=E_{\varphi}^{(1)}=E_{\varphi}^{(2)}=0,
\end{equation}
and all of the magnetic fields are in z-direction i.e. they are
perpendicular to $r,\phi$ plane. These magnetic fields are given
by
\begin{equation}\label{L30}
B_{z}^{(0)}=-\frac{J^2}{2r^4}h(r)+\frac{N^2}{r}\frac{d}{dr}h(r),~~~~~~~~~~ B_{z}^{(1)}=\frac{2k(M+\Lambda r^2)}{g(r)}+\frac{kr
N^2}{(g(r))^2}\frac{d}{dr}g(r),~~~~~~~~~~ B_{z}^{(2)}=0.~~~~
\end{equation}

\section {\large {\bf Dual solution}}

Now, we will try to find the dual of the Abelian solution
\eqref{L19} which is indeed another solution for the equations of motion.
Note that the solution \eqref{L19} is independent of the coordinate
$\varphi$; hence we have an isometry in $\varphi$ direction.
Abelian duality which is a transformation on the string model, relates
this solution $(g_{\mu\nu}, B_{\mu\nu}, F^{~~a}_{\mu\nu}, \phi)$
to the dual solution $(\tilde{g}_{\mu\nu}, \tilde{B}_{\mu\nu},
\tilde{F}^{~~a}_{\mu\nu}, \tilde{\phi})$ by the following
Buscher's transformation \cite{Buscher}

\begin{equation}\nonumber
\tilde{g}_{22}=\frac{1}{g_{22}},~~~~~~~~~~~~~~~~~~~~~~~~~~
\tilde{g}_{\alpha\beta}=g_{\alpha\beta}-\frac{(g_{2\alpha}g_{2\beta}
-B^{\prime}_{2\alpha}B^{\prime}_{2\beta})}{g_{22}},~~~
\end{equation}
\begin{equation}\label{L31}
\tilde{g}_{2\alpha}=\frac{B^{\prime}_{2\alpha}}{g_{22}},~~~~~~~~~~~~~~~~~~~~~~~
\tilde{B^{\prime}}_{\alpha\beta} = B^{\prime}_{\alpha\beta}-\frac{2~g_{2[\alpha} ~B^{\prime}_{\beta]2} }{g_{22}},~~~~~~~~~~~~~
\end{equation}
\begin{equation}\nonumber
\tilde{B^{\prime}}_{2\alpha} = \frac{g_{2\alpha}}{g_{22}},~~~~~~~~~~~~~~~~~~~~~~~~~~~~
\tilde{\phi}=\phi-\frac{1}{2}ln(g_{22}),~~~~~~~~~~~~~~~~~~~~~~
\end{equation}
where $\alpha, \beta=0,1$ (indice `$2$' represents the $\varphi$
coordinate). Applying this transformation to the solution
\eqref{L11},\eqref{L12},\eqref{L19} and \eqref{L21} yields the
following dual solution:
\begin{equation}\nonumber
\tilde{ds}^{2}=\Big(M-\frac{J^2}{4r^2}\Big)dt^{2}+\frac{2}{\ell}dtd\varphi
+2 B^{\prime}_{21}(r) \Big(\frac{1}{\ell}dtdr
+\frac{1}{r^2}drd\varphi\Big)
+\frac{1}{r^2}d\varphi^{2}+\Big\{\frac{1}{N^{2}(r)}+\frac{1}{r^2}\Big(B^{\prime}_{21}(r)\Big)^{2}\Big\}dr^2,
\end{equation}
\begin{equation}\nonumber
\tilde{B}_{20}(r)=-\frac{J}{2r^2},~~~~~~~~~~~
\tilde{\phi}(r)=-ln(r),~~~~~~~~~~~~~~~~~~~~~~~~~~~~~~~~~~~~~~~~~~~~~~~~~~~~~~~~~~~~~~~~~~~~~~~~~~~~~~~~~~~~~~~~~
\end{equation}
\begin{equation}\label{L32}
\tilde{F}^{~~0}_{01}(r)=-\frac{J}{r^3}h(r)+\frac{3J}{2r^2}\frac{d}{dr}h(r),~~~~~~~~~~~~~
\tilde{F}^{~~1}_{01}(r)=-\frac{2Jk}{r g(r)}+
\frac{3Jk}{2 g^2(r)}\frac{d}{dr}g(r),~~~~~~~~~~~~~~~~~~~~~~~~~~~~~~~~~~~~~~~
\end{equation}
where the dual electromagnetic field strengths
$\tilde{F}^{~~0}_{01}$ and $\tilde{F}^{~~1}_{01}$, are related to
the following gauge fields (using \eqref{L20}):
\begin{equation}\nonumber
\tilde{A}_{0}^{~0}(r)=-\int dr
(-\frac{J}{r^3}h(r)+\frac{3J}{2r^2}\frac{d}{dr}h(r)),~~~~~~
\end{equation}
\begin{equation}\label{L33}
\tilde{A}_{0}^{~1}(r)=-\int dr (-\frac{2Jk}{r g(r)}+
\frac{3Jk}{2 g^2(r)}\frac{d}{dr}g(r)),~~
\end{equation}
\begin{equation}\nonumber
\tilde{A}_{1}^{~2}(r)=\frac{-1}{N^{2}(r)}\int dr
(-\frac{J}{r^3}h(r)+\frac{3J}{2r^2}\frac{d}{dr}h(r)).
\end{equation}
Note that for $B^{\prime}_{21}(r)=0$, and using the following coordinate
transformation:
\begin{equation}\label{L34}
r^2=\ell r^{\prime},~~~~~~~~~~~~~~~ t=\frac{\sqrt{\ell}
(\varphi^{\prime}-t^{\prime})}{(M^{2}\ell^2-J^{2})^{\frac{1}{4}}},~~~~~~~~~~~~~~
\varphi=\frac{r_{+}^{2} t^{\prime}-r_{-}^{2}
\varphi^{\prime}}{\sqrt{\ell}(M^{2}\ell^2-J^{2})^{\frac{1}{4}}},
\end{equation}
the dual metric in \eqref{L32} precisely represents the charged black string
solution \cite{Horne,Horowitz}.
The dual electric and magnetic fields can be obtained in terms of
dual field strengths as follows:
\begin{equation}\nonumber
\tilde{E}_{r}^{(a)}=-\tilde{F}^{01a}=-\tilde{g}^{00}\tilde{g}^{11}\tilde{F}_{01}^{~~a},~~~~~~~~~~~
\tilde{E}_{\varphi}^{(a)}=-\tilde{F}^{02a}=-\tilde{g}^{00}\tilde{g}^{21}\tilde{F}_{01}^{~~a},
\end{equation}
\begin{equation}\label{L35}
\tilde{B}_{z}^{(a)}=r\tilde{F}^{12a}=-r\tilde{g}^{11}\tilde{g}^{20}\tilde{F}_{01}^{~~a}
= -\frac{r^3}{\ell} \tilde{F}_{01}^{~~a}.~~~~~~~~~~~~~~~~~~~~~~~~~~~~~~~~~
\end{equation}
Furthermore, the radial components of three dual electric fields
are obtained as:
\begin{equation}\label{L36}
\tilde{E}_{r}^{(0)}=-\frac{J}{r^3}h(r)+\frac{3J}{2r^2}\frac{d}{dr}h(r),~~~~~~~~~~
\tilde{E}_{r}^{(1)}=-\frac{2Jk}{r g(r)}+
\frac{3Jk}{2 g^2(r)}\frac{d}{dr}g(r),~~~~~~~~~~
\tilde{E}_{r}^{(2)}=0,
\end{equation}
and the azimuthal components of them are obtained as follows:
\begin{equation}\nonumber
\tilde{E}_{\varphi}^{(0)}=\Big(\frac{d}{dr}h(r)-\frac{2kr}{g(r)}+\frac{k
r^2}{g^2(r)}\frac{d}{dr}g(r)\Big)~ \Big(
-\frac{J}{r^3}h(r)+\frac{3J}{2r^2}\frac{d}{dr}h(r) \Big),~~~~
\end{equation}
\begin{equation}\label{L37}
\tilde{E}_{\varphi}^{(1)}
=\Big(\frac{d}{dr}h(r)-\frac{2kr}{g(r)}+\frac{k
r^2}{g^2(r)}\frac{d}{dr}g(r)\Big)~ \Big( -\frac{2Jk}{r g(r)}+
\frac{3Jk}{2 g^2(r)}\frac{d}{dr}g(r) \Big),~~
\end{equation}
\begin{equation}\nonumber
\tilde{E}_{\varphi}^{(2)}=0,~~~~~~~~~~~~~~~~~~~~~~~~~~~~~~~~~~~~~~~~~~~~~~~~~~~~~~~~~~~~~~~~~~~~~~~~~~~~~~~~
\end{equation}
and finally the magnetic fields which all are in z-direction are
given by the following relations:
\begin{equation}\label{L38}
\tilde{B}_{z}^{(0)}=-\frac{r^3}{\ell} \Big(
-\frac{J}{r^3}h(r)+\frac{3J}{2r^2}\frac{d}{dr}h(r) \Big),~~~~~~~~
\tilde{B}_{z}^{(1)}=-\frac{r^3}{\ell} \Big( -\frac{2Jk}{r g(r)}+ \frac{3Jk}{2 g^2(r)}\frac{d}{dr}g(r) \Big),~~~~~~~~
\tilde{B}_{z}^{(2)}=0.
\end{equation}
Note that if we select the following forms for arbitrary functions
$g(r)$ and $h(r)$:
\begin{equation}\label{L39}
g(r)=D r N(r)~e^{-V(r)},~~~~~~~~~~ h(r)=\frac{C r}{N(r)}
~e^{V(r)},
\end{equation}
where $D$ and $C$ are arbitrary constants and
\begin{equation}\label{L40}
V(r)= \frac{M}{\sqrt{M^2+\Lambda J^2}}~
tanh^{-1}(\frac{M+2\Lambda r^2}{\sqrt{M^2+\Lambda J^2}}),
\end{equation}
then all magnetic fields in \eqref{L30} become zero, and
therefore, we have only the electric fields in solution
\eqref{L19}, while in dual solution \eqref{L32} there are both
electric and magnetic fields. This result, as we expect, indicates that here, in
string theory, the duality leads to a connection between
electric and magnetic fields in solution \eqref{L19} and it's
dual solution \eqref{L32}.

At the end of this section let us discuss about the isometry symmetry along $t$ direction of the solution \eqref{L19}. One can repeat the above procedure to
find the following $t$-dual solution \cite{Welch} for solution \eqref{L19}:
\begin{equation}\nonumber
\tilde{ds}^{2}=\frac{1}{(M-\frac{r^2}{\ell^2})} \Big[dt^{2}+2 \frac{~r^2}{\ell}dtd\varphi
-2 B^{\prime}_{01}(r)\Big(dtdr-\frac{~r^2}{\ell}drd\varphi\Big)
+r^2(M-\frac{J^2}{4r^2})d\varphi^{2} +\Big\{\frac{(M-\frac{r^2}{\ell^2})}{N^{2}(r)}+\Big(B^{\prime}_{01}(r)\Big)^{2}\Big\}dr^2\Big],
\end{equation}
\begin{equation}\nonumber
~\tilde{B}_{20}(r)=-\frac{J}{2(M-\frac{r^2}{\ell^2})},~~~~~~~~~~~~
\tilde{\phi}(r)=-\frac{1}{2}ln(-M+\frac{r^2}{\ell^2}),~~~~~~~~~~~~~~~~~~~~~~~~~~~~~~~~~~~~~~~~~~~~~~~~~~~~~~~~~~~~~~~~~~~~~~~~~~~~~~~~~~
\end{equation}
\begin{equation}\label{L41}
\tilde{F}^{~~0}_{21}(r)= \frac{1}{(M-\frac{r^2}{\ell^2})} \Big( \frac{J^2}{r^3}h(r)+(-M+\frac{r^2}{\ell^2}-\frac{J^2}{2r^2})\frac{d}{dr}h(r) \Big),~~~~~~~~~~~~~~~~~~~~~~~~~~~~~~~~~~~~~~~~~~~~~~~~~~~~~~~~~~~~~~~~~~~~~~~~~~~~~~~~~~~~~~~~~~~~~~~~~~~
\end{equation}
\begin{equation}\nonumber
\tilde{F}^{~~1}_{21}(r)=\frac{2kr}{g(r)}+
\frac{kr^2}{g^2(r)(M-\frac{r^2}{\ell^2})}(-M+\frac{r^2}{\ell^2}-\frac{J^2}{2r^2})\frac{d}{dr}g(r),~~~~~~~~~~~~~~~~~~~~~~~~~~~~~~~~~~~~~~~~~~~~~~~~~~~~~~~~~~~~~~~~~~~~~~~~~~~~~~~~~~~~~~~~~~~~~~~
\end{equation}
For $B^{\prime}_{01}(r)=0,$ and using the following coordinate
transformation \cite{Welch}:
\begin{equation}\label{L42}
r^2=\ell r^{\prime}+M \ell^2,~~~~~~~~~~~~~~~ t=\frac{r_{+}^{2} \varphi^{\prime}-r_{-}^{2}
 t^{\prime}}{\ell^{\frac{3}{2}}(M^{2}\ell^2-J^{2})^{\frac{1}{4}}}~~~~~~~~~~~~~~
\varphi=\frac{
t^{\prime}-\varphi^{\prime}}{\sqrt{\ell}(M^{2}\ell^2-J^{2})^{\frac{1}{4}}},
\end{equation}
the $t$-dual metric in \eqref{L41} precisely reduces to the charged black string solution.
Finally, calculations similar to the $\varphi$-dual case show that $t$-dual solution also leads
to a connection between electric and dual magnetic fields for solution
\eqref{L19} and it's $t$-dual solution, respectively.

\section {\large {\bf Conclusions}}

We have presented a $2+1$ dimensional low energy string effective
action containing gauge fields term which has a gauge symmetry
coming from semi-simple extension of Poincar\'{e} (Maxwell) gauge
group. The model has led to an extended B-field in the
corresponding sigma model. By solving the equations of motion of
the string effective action (i.e. the beta function equations),
we have obtained two different solutions which both in the sigma
model level correspond to the $SL(2,R)$ WZW models and then, both
are exact solutions of beta function equations to all orders.
Also, it turned out that two sigma models corresponding to two
different solutions are classically canonically equivalent. We
have interpreted the gauge field strength tensors related to the
Abelian gauge fields solution as electromagnetic field strength
tensors and obtained the corresponding electric and magnetic
fields. Using Buscher duality transformation, we have shown that
the dual models coincide with the charged black string solution
and also have shown that the electric fields of the Abelian
solution are related to the magnetic fields of it's dual solution.

\section*{Acknowledgements}
We would like to express our gratitude to M.M. Sheikh-Jabbari and F. Darabi for their useful comments.
This research was supported by a research fund No. 217D4310 from Azarbaijan Shahid Madani university.

\appendix

\section{\large {\bf Appendix: S-expansion of the anti-de Sitter ($Ads$) algebra $so(2,2)$ in $2+1$ dimensions}}
\setlength{\parindent}{0cm}

In this section, we start from the anti-de Sitter algebra
$\mathfrak{g}=so(2,2)$ in $2+1$ dimensions and use the finite
abelian semigroup expansion procedure\footnote{The full details
of the S-expansion procedure have been presented in
\cite{Izaurieta-1}.} (S-expansion) to cast both the Maxwell
algebra and the semi-simple extension of the Poincar\'{e}
algebra.\footnote{The Maxwell algebra and the semi-simple
extension of the Poincar\'{e} algebra in $D$-dimensional
spacetime have been obtained by the S-expansion of the
$D$-dimensional $Ads$ algebra in \cite{Salgado} and \cite{Diaz},
respectively.} We consider the anti-de Sitter algebra
$\mathfrak{g}=so(2,2)$ with the basis
$\overline{X}_{B}=\{\overline{P}_{a},\overline{J}_{a}\}$ in $2+1$
dimensions as follows:
\begin{equation}   \label{L43}
[\overline{J}_{a},\overline{J}_{b}] = \epsilon_{abc}
\overline{J}^{c}, ~~~~~~~[\overline{J}_{a},\overline{P}_{b}] =
\epsilon_{abc} \overline{P}^{c},
~~~~~~~~~~[\overline{P}_{a},\overline{P}_{b}] = \Lambda
\epsilon_{abc} \overline{J}^{c},
\end{equation}
where $\Lambda$ is a constant, and $\overline{P}_{a}$ and
$\overline{J}_{a}$ are the ordinary translation and Lorentz
generators, respectively. We split the $Ads$ algebra $so(2,2)$ in
two subspaces $so(2,2)=V_{0}\bigoplus V_{1}$, where $V_{0}$ and
$V_{1}$ corresponds to the Lorentz and translation generators
$\overline{J}_{a}$ and $\overline{P}_{a}$, respectively. The
subspace structure is such that we have:
\begin{equation}   \label{L44}
[V_{0},V_{0}] \subset V_{0}, ~~~~~~~[V_{0},V_{1}] \subset V_{1},
~~~~~~~~[V_{1},V_{1}] \subset V_{0}.
\end{equation}
Now, in the following, we use two different semigroups to expand
the $Ads$ algebra by use of the S-expansion procedure, and obtain
both the Maxwell algebra and the semi-simple extension of the
Poincar\'{e} algebra.

\subsection {\large {\bf The Maxwell algebra and the S-expansion by the semigroup $S$}}

We first consider the abelian semigroup
$S=\{\lambda_{0},\lambda_{1},\lambda_{2},\lambda_{3}\}$ together
with the following multiplication law:
\begin{equation}   \label{L45}
\lambda_{\alpha}\lambda_{\beta}=\Big\{
\begin{array}{cc}
  \lambda_{3} & ~if~~(\alpha+\beta)>2 \\
  \lambda_{\alpha+\beta} & ~if~~(\alpha+\beta)\leq 2
\end{array},
\end{equation}
or, equivalently, by the following multiplication table:
\begin{equation}   \label{L46}
\begin{array}{c|cccc}
              & \lambda_{0} & \lambda_{1} & \lambda_{2} & \lambda_{3} \\ \hline
  \lambda_{0} & \lambda_{0} & \lambda_{1} & \lambda_{2} & \lambda_{3} \\
  \lambda_{1} & \lambda_{1} & \lambda_{2} & \lambda_{3} & \lambda_{3} \\
  \lambda_{2} & \lambda_{2} & \lambda_{3} & \lambda_{3} & \lambda_{3} \\
  \lambda_{3} & \lambda_{3} & \lambda_{3} & \lambda_{3} & \lambda_{3}
\end{array}
\end{equation}
Note that for each $\lambda_{\alpha}\in S$, we have
$\lambda_{3}\lambda_{\alpha}=\lambda_{3}$, such that
$\lambda_{3}$ plays the role of the zero element inside the
semigroup $S$ (i.e. $\lambda_{3}=0_{S}$). Now, we consider the
following subset decomposition $S=S_{0}\bigcup S_{1}$:
\begin{equation}   \label{L47}
S_{0}=\{\lambda_{0},\lambda_{2},\lambda_{3}\},~~~~~~~~
S_{1}=\{\lambda_{1},\lambda_{3}\},
\end{equation}
which is said to be a resonant decomposition, in other words, it
is in resonance with the subspace decomposition $\mathfrak{g}=V_{0}\bigoplus
V_{1}$ and then, satisfies the following resonance condition:
\begin{equation}   \label{L48}
S_{0}.S_{0} \subset S_{0}, ~~~~~~~S_{0}.S_{1} \subset S_{1},
~~~~~~~~S_{1}.S_{1} \subset S_{0}.
\end{equation}
The direct product $S\times \mathfrak{g}$ with basis $\lambda_{\alpha}
\overline{X}_{B}$ is a Lie algebra (see Theorem III.1 in
\cite{Izaurieta-1}). According to the Theorem IV.2 in
\cite{Izaurieta-1}, $W_{0}\oplus W_{1}$ is a resonant subalgebra
of $S\times \mathfrak{g}$ where we have:
\begin{equation}   \label{L49}
W_{0}=S_{0}\times V_{0}
=\{\lambda_{0},\lambda_{2},\lambda_{3}\}\otimes
\{\overline{J}_{a}\}
=\{\lambda_{0}\overline{J}_{a},\lambda_{2}\overline{J}_{a},\lambda_{3}\overline{J}_{a}\},
\end{equation}
\begin{equation}   \label{L50}
W_{1}=S_{1}\times V_{1} =\{\lambda_{1},\lambda_{3}\}\otimes
\{\overline{P}_{a}\}
=\{\lambda_{1}\overline{P}_{a},\lambda_{3}\overline{P}_{a}\}.
\end{equation}
Now, we impose the condition $\lambda_{3}\times \mathfrak{g}=0_{S}$, and
remove the whole $0_{S}\times \mathfrak{g}$ sector from the resonant
subalgebra. The remaining piece is a Lie algebra and is called the
$0_{S}$-reduced algebra (see $0_{S}$-reduction and Theorem VI.1 in
\cite{Izaurieta-1}). By relabeling the generators as
$J_{a}\equiv\lambda_{0}\overline{J}_{a},
~kZ_{a}\equiv\lambda_{2}\overline{J}_{a}$ and
$\sqrt{\Lambda}P_{a}\equiv\lambda_{1}\overline{P}_{a}$, we obtain
the following commutation relations:
\begin{equation}   \nonumber
[J_{a},J_{b}]=\lambda_{0}\lambda_{0}[\overline{J}_{a},\overline{J}_{b}]
=\lambda_{0}\epsilon_{abc} \overline{J}^{c} =\epsilon_{abc} J^{c},
\end{equation}
\begin{equation}   \nonumber
[J_{a},P_{b}]=\frac{1}{\sqrt{\Lambda}}\lambda_{0}\lambda_{1}[\overline{J}_{a},\overline{P}_{b}]
=\frac{1}{\sqrt{\Lambda}}\lambda_{1}\epsilon_{abc}
\overline{P}^{c} =\epsilon_{abc} P^{c},
\end{equation}
\begin{equation}   \nonumber
[P_{a},P_{b}]=\frac{1}{\Lambda}\lambda_{1}\lambda_{1}[\overline{P}_{a},\overline{P}_{b}]
=\lambda_{2}\epsilon_{abc} \overline{J}^{c} =k\epsilon_{abc}
Z^{c},
\end{equation}
\begin{equation}   \label{L51}
[J_{a},Z_{b}]=\frac{1}{k}\lambda_{0}\lambda_{2}[\overline{J}_{a},\overline{J}_{b}]
=\frac{1}{k}\lambda_{2}\epsilon_{abc} \overline{J}^{c}
=\epsilon_{abc} Z^{c},
\end{equation}
\begin{equation}   \nonumber
[P_{a},Z_{b}]=\frac{1}{k\sqrt{\Lambda}}\lambda_{1}\lambda_{2}[\overline{P}_{a},\overline{J}_{b}]
=\frac{1}{k\sqrt{\Lambda}}\epsilon_{abc}
\lambda_{3}\overline{P}^{c}=0,
\end{equation}
\begin{equation}   \nonumber
[Z_{a},Z_{b}]=\frac{1}{k^2}\lambda_{2}\lambda_{2}[\overline{J}_{a},\overline{J}_{b}]
=\frac{1}{k^2}\epsilon_{abc} \lambda_{3}\overline{J}^{c}=0,
\end{equation}
where we have used the commutation relations of the $Ads$ algebra
\eqref{L43} and the multiplication law \eqref{L46} of the semigroup
$S$. The obtained algebra \eqref{L51} coincides with the Maxwell
algebra \cite{Bacry,Schrader}.

\subsection {\large {\bf The semi-simple extension of the Poincar\'{e} algebra and the S-expansion by the semigroup
$\overline{S}$}}

We consider the abelian semigroup
$\overline{S}=\{\overline{\lambda}_{0},\overline{\lambda}_{1},\overline{\lambda}_{2}\}$
together with the following multiplication law:
\begin{equation}   \label{L52}
\overline{\lambda}_{\alpha}\overline{\lambda}_{\beta}=\Big\{
\begin{array}{cc}
  \overline{\lambda}_{\alpha+\beta-2} & ~if~~(\alpha+\beta)>2 \\
  \overline{\lambda}_{\alpha+\beta} & ~if~~(\alpha+\beta)\leq 2
\end{array},
\end{equation}
or, equivalently, by the following multiplication table:
\begin{equation}   \label{L53}
\begin{array}{c|cccc}
                         & \overline{\lambda}_{0} & \overline{\lambda}_{1} & \overline{\lambda}_{2} \\ \hline
  \overline{\lambda}_{0} & \overline{\lambda}_{0} & \overline{\lambda}_{1} & \overline{\lambda}_{2} \\
  \overline{\lambda}_{1} & \overline{\lambda}_{1} & \overline{\lambda}_{2} & \overline{\lambda}_{1} \\
  \overline{\lambda}_{2} & \overline{\lambda}_{2} & \overline{\lambda}_{1} & \overline{\lambda}_{2} \\
\end{array}
\end{equation}
Now, we consider the following subset decomposition
$\overline{S}=\overline{S}_{0}\bigcup \overline{S}_{1}$:
\begin{equation}   \label{L54}
\overline{S}_{0}=\{\overline{\lambda}_{0},\overline{\lambda}_{2}\},~~~~~~~~
\overline{S}_{1}=\{\overline{\lambda}_{1}\},
\end{equation}
which is in resonance with the subspace decomposition
$\mathfrak{g}=V_{0}\bigoplus V_{1}$ (resonant decomposition) and then,
satisfies the following resonance condition:
\begin{equation}   \label{L55}
\overline{S}_{0}.\overline{S}_{0} \subset \overline{S}_{0},
~~~~~~~ \overline{S}_{0}.\overline{S}_{1} \subset
\overline{S}_{1}, ~~~~~~~~ \overline{S}_{1}.\overline{S}_{1}
\subset \overline{S}_{0}.
\end{equation}
The direct product $\overline{S}\times \mathfrak{g}$ with basis
$\overline{\lambda}_{\alpha} \overline{X}_{B}$ is a Lie algebra
(see Theorem III.1 in \cite{Izaurieta-1}). According to the
Theorem IV.2 in \cite{Izaurieta-1}, $\overline{W}_{0}\oplus
\overline{W}_{1}$ is a resonant subalgebra of $\overline{S}\times
\mathfrak{g}$ where we have:
\begin{equation}   \nonumber
\overline{W}_{0}=\overline{S}_{0}\times V_{0}
=\{\overline{\lambda}_{0},\overline{\lambda}_{2}\}\otimes
\{\overline{J}_{a}\}
=\{\overline{\lambda}_{0}\overline{J}_{a},\overline{\lambda}_{2}\overline{J}_{a}\},
\end{equation}
\begin{equation}   \label{L56}
\overline{W}_{1}=\overline{S}_{1}\times V_{1}
=\{\overline{\lambda}_{1}\}\otimes \{\overline{P}_{a}\}
=\{\overline{\lambda}_{1}\overline{P}_{a}\}.
\end{equation}
Relabeling the generators as
$J_{a}\equiv\overline{\lambda}_{0}\overline{J}_{a},
~-\frac{k}{\lambda}Z_{a}\equiv\overline{\lambda}_{2}\overline{J}_{a}$
and
$\sqrt{-\frac{\Lambda}{\lambda}}P_{a}\equiv\overline{\lambda}_{1}\overline{P}_{a}$,
we obtain the following commutation relations:
\begin{equation}   \nonumber
[J_{a},J_{b}]=\overline{\lambda}_{0}\overline{\lambda}_{0}[\overline{J}_{a},\overline{J}_{b}]
=\overline{\lambda}_{0}\epsilon_{abc} \overline{J}^{c}
=\epsilon_{abc} J^{c},
\end{equation}
\begin{equation}   \nonumber
[J_{a},P_{b}]=\sqrt{-\frac{\lambda}{\Lambda}}~\overline{\lambda}_{0}\overline{\lambda}_{1}[\overline{J}_{a},\overline{P}_{b}]
=\sqrt{-\frac{\lambda}{\Lambda}}~\overline{\lambda}_{1}\epsilon_{abc}
\overline{P}^{c} =\epsilon_{abc} P^{c},
\end{equation}
\begin{equation}   \nonumber
[P_{a},P_{b}]=-\frac{\lambda}{\Lambda}\overline{\lambda}_{1}\overline{\lambda}_{1}[\overline{P}_{a},\overline{P}_{b}]
=-\lambda \overline{\lambda}_{2}\epsilon_{abc} \overline{J}^{c}
=k\epsilon_{abc} Z^{c},
\end{equation}
\begin{equation}   \label{L57}
[J_{a},Z_{b}]=-\frac{\lambda}{k}\overline{\lambda}_{0}\overline{\lambda}_{2}[\overline{J}_{a},\overline{J}_{b}]
=-\frac{\lambda}{k}\overline{\lambda}_{2}\epsilon_{abc}
\overline{J}^{c} =\epsilon_{abc} Z^{c},
\end{equation}\begin{equation}   \nonumber
[P_{a},Z_{b}]=-\frac{\lambda}{k}\sqrt{-\frac{\lambda}{\Lambda}}
~\overline{\lambda}_{1}\overline{\lambda}_{2}[\overline{P}_{a},\overline{J}_{b}]
=-\frac{\lambda}{k}\sqrt{-\frac{\lambda}{\Lambda}}
~\overline{\lambda}_{1} \epsilon_{abc}
\overline{P}^{c}=-\frac{\lambda}{k}\epsilon_{abc}\overline{P}^{c},
\end{equation}
\begin{equation}   \nonumber
[Z_{a},Z_{b}]=\frac{\lambda^2}{k^2}\overline{\lambda}_{2}\overline{\lambda}_{2}[\overline{J}_{a},\overline{J}_{b}]
=\frac{\lambda^2}{k^2}\overline{\lambda}_{2}\epsilon_{abc}\overline{J}^{c}
=-\frac{\lambda}{k}\epsilon_{abc}\overline{Z}^{c},
\end{equation}
where we have used the commutation relations of the $Ads$ algebra
\eqref{L43} and the multiplication law \eqref{L53} of the semigroup
$\overline{S}$. The obtained algebra \eqref{L57} matches the
semi-simple extension of the Poincar\'{e} algebra \cite{Soroka}.


\end{document}